\documentstyle[11pt,aaspp4]{article}

\def\la{\mathrel{\hbox{\rlap{\hbox{\lower4pt\hbox{$\sim$}}}\hbox{$<$}}}}
\def\ga{\mathrel{\hbox{\rlap{\hbox{\lower4pt\hbox{$\sim$}}}\hbox{$>$}}}}

\slugcomment{Submitted to {\it The Astrophysical Journal Letters}}

\begin{document}

\title{Observations and Theoretical Implications of GRB 970228}

\author{Daniel E. Reichart}

\affil{Department of Astronomy and Astrophysics, University of Chicago, Chicago, IL 60637} 

\begin{abstract}

GRB 970228 is the first gamma ray burst for which prolonged post-burst transient x-ray, optical, and infrared emission has been detected.  Recent Hubble Space Telescope observations show that the transient consists of two components:  a point source, which is known to be fading, and an extended source, which is possibly fading.  I fit standard fireball remnant models to the first month of x-ray, optical, and infrared measurements, which may be done without assuming a GRB distance scale.  
I show that its emission is consistent with that of the remnant of a relativistically expanding impulsive fireball in which a forward shock dominates the emission of the GRB event:  the piston model.  
However, two discrepant measurements may indicate that the post-burst flux varies by factors of $\sim$ 3 on timescales of days or weeks.  
Furthermore, using the HST observations and the fitted model, I show that the extended object probably is fading, which may place GRB 970228 at galactic halo distances.  

\end{abstract}


\section{Introduction}

Discovered by the BeppoSAX Gamma Ray Burst Monitor (Costa et al. 1997a), GRB 970228 is the first gamma ray burst (GRB) for which non-gamma ray emission has been detected for a prolonged period of time after a gamma ray burst:  Costa et al. (1997b), Yoshida et al. (1997), and Frontera et al. (1997) report transient x-ray emission from $\sim$ 8 hours to $\sim$ 13 days after the GRB event; Groot et al. (1997a,b), Metzger et al. (1997a,b), Sahu et al. (1997a,b), Margon et al. (1997), van Paradijs et al. (1997), and Pedichini et al. (1997) report transient optical emission from $\sim$ 17 hours to $\sim$ 37 days after the GRB event; and Klose et al. (1997) and Soifer et al. (1997) report transient infrared emission from $\sim$ 17 days to $\sim$ 30 days after the GRB event.  
Previously, only x-ray emission has been detected after GRB events, and then, for no longer than several hundred seconds (Murakami et al. 1992).  Optical and infrared emission have never been detected either during or after a GRB event, making this the first GRB optical/infrared counterpart ever detected.  
In Table 1, I list all detections and upper bounds reported before 1997 May 11.

Additionally, Groot et al. (1997b) and Metzger et al. (1997a) report the existence of an extended object at the position of the previously reported optical transient.  
Furthermore, Sahu et al. (1997a,b), using the Hubble Space Telescope Wide Field and Planetary Camera, report the existence of a fading point source embedded in an extended object.  
If this object is a galaxy, and unless the point source is coincident by chance, this would mark the first identification of a GRB host, and it would place the bursts at cosmological distances.  
However, a recent report that the extended object is fading (Metzger et al. 1997b) demonstrates that the GRB distance scale is still an open question.  

HST images (V- and I-band) were taken on March 26 and on April 7 (Sahu et al 1997a,b).  The March 26 HST images suggest that the total emission of the extended object is comparable to that of the point source on this date.  
Consequently, I consider three scenarios:  (1) the extended object is not fading, in which case all detections before March 26 are dominated by the emission from the point source and all later detections are dominated by the emission from the extended object; (2) the extended object is fading, but the point source has always dominated the emission; and (3) the extended source is fading, and at some time before March 26, it dominated the emission.  
In \S2, the first two scenarios are examined in terms of fireball models.  In \S3, the third scenario is discussed.  Conclusions are drawn in \S4.

\section{Fireball Models And The Dominant Point Source Scenario}

GRBs, whether at cosmological or galactic halo distances, have been theorized to be caused by relativistically expanding fireballs that dissipate their energy after becoming optically thin.  This occurs in shocks, which are produced either through interaction with an external medium (Rees \& M\'esz\'aros 1992, M\'esz\'aros \& Rees 1993, Katz 1994, M\'esz\'aros, Rees, \& Papathanassiou 1994, Sari, Narayan, \& Piran 1996), or internally (Rees \& M\'esz\'aros 1994, Paczy\'nski \& Xu 1994, Papathanassiou \& M\'esz\'aros 1996).  In the former case, the initial energy input is impulsive, and both forward and reverse shocks are possible.  In the latter case, the initial energy input is prolonged, resulting in a relativistic wind in which internal shocks dissipate the bulk of the energy before interaction with an external medium becomes important.
Both impulsive and wind fireballs are predicted to leave behind expanding, cooling, GRB remnants (GRBRs), the emission from which should be detectable at x-ray and optical frequencies for hours to days after a GRB event (M\'esz\'aros \& Rees 1997).  
Furthermore, standard types of GRBRs (i.e., forward shocking impulsive fireball GRBRs, reverse shocking impulsive fireball GRBRs, etc.) can be characterized by power-law spectra and by power-law temporal decays for the frequencies in question (M\'esz\'aros \& Rees 1997, Papathanassiou \& M\'esz\'aros 1996, Rees, M\'esz\'aros, \& Papathanassiou 1994).  
Consequently, to determine if the fading GRB 970228 point source can be described by one or more existing GRBR models, I fit the following simple power law form to the reported flux (and magnitude) measurements given in Table 1:
\begin{equation}
F_\nu = F_0\nu^{-a}t^{-b},
\end{equation}
where $F_0$, $a$, and $b$ are free parameters, and $F_\nu$ is measured in erg cm$^{-2}$ s$^{-1}$ Hz$^{-1}$.  A spectral index of $a \sim$ 0.5 - 1.5 is expected, due to synchrotron and/or inverse Compton radiation (Rees, M\'esz\'aros, \& Papathanassiou 1994); however, the values of $b$ and $F_0$ can be used to discriminate between existing GRBR models (M\'esz\'aros \& Rees 1997).  I assume neither a cosmological nor a galactic halo distance scale.

\subsection{Data Analysis}

In all, 25 measurements of the GRB transient have been reported:  7 in $\sim$ two x-ray bands, 15 in $\sim$ four optical bands, and 3 in two infrared bands.  
Seven of these measurements, I exclude from the fit:  the three x-ray measurements of Yoshida et al. (1997) and Frontera et al. (1997) depend on an assumed spectral form, the optical measurement of Pedichini et al. (1997) is broadband, and the three ground-based optical and infrared measurements (R-, J-, and K-bands) taken after March 26 may be dominated by the emission of the extended object, in the case that it is not fading (\S1).  However, I find that all seven excluded measurements are consistent with the best fit.  
The 18 optical and infrared measurements (and 6 optical and infrared upper bounds) are corrected for galactic extinction using the hydrogen column densities of Stark et al. (1992).  
Due to the low galactic latitude of GRB 970228, corrections can be as large as 1.1 magnitudes (B-band).  In the x-ray and gamma ray bands, corrections are expected to be $\sim$ 0.1 magnitudes (0.5 - 2 keV) and $\sim$ 0.0 magnitudes (2 - 10 keV, 40 - 80 keV), and are consequently ignored.  

I fit Equation 1 to the remaining 18 measurements, and find the following best-fit parameters values:  $\log F_0 = -9.7 \pm 2.8$, $a = 0.86 \pm 0.12$, and $b = 1.09 \pm 0.23$. 
The standard deviation about the best fit is $\pm 0.65$ magnitudes, which is approximately twice as large as what one would expect from photometric errors alone.  However, if the J-band measurement of March 17 (Klose et al. 1997), which is 1.3 magnitudes brighter than the best fit (\S2.2), is ignored, the standard deviation of the remaining fitted measurements is $\pm 0.47$ magnitudes.  Given the accuracy to which the extinction can be corrected, this is consistent with the expected standard photometric error of $\sim$ $\pm 0.3$ magnitudes.
The 1-$\sigma$ errors of the best fit assume a constant standard error equal to the fitted standard deviation, $\pm 0.65$ magnitudes, for each of the 18 fluxes fitted to, and are consequently only approximate.  
The best fit and the extinction corrected fluxes and flux upper bounds are plotted in Figure 1.  The best fit spectral form has been assumed for those x-ray and gamma ray fluxes that require the spectral form to be specified.  
These fluxes are also consistent with the best fit to within the expected photometric error.

\subsection{Discussion}

The fitted value of $a = 0.86 \pm 0.12$ is consistent with the expected value, 0.5 $\la a \la$ 1.5, but it does not discriminate between the standard GRBR models.  
The fitted value of $b = 1.09 \pm 0.23$, however, is highly discriminatory:  of the standard GRBR models, only an impulsive fireball in which a forward shock dominates the emission exhibits a similar temporal decay:  $b \sim$ 1.5 (M\'esz\'aros \& Rees 1997).  
Impulsive fireballs in which a reverse shock dominates the emission have $b \sim 2$ and wind fireballs have $b \ga 6$ (M\'esz\'aros \& Rees 1997).  
For the forward shocking impulsive fireball GRBR, also called the piston model, M\'esz\'aros \& Rees (1997) additionally estimate $F_0$:
\begin{equation}
F_0 \sim 10^{-7} E_{51} \theta_{-1}^{-2} D_{28}^{-2} t_{\gamma}^{1.5} = 10^{-7} E_{43} \theta_{-1}^{-2} D_{24}^{-2} t_{\gamma}^{1.5},
\end{equation}
where $10^{51} E_{51}$ erg (or $10^{43} E_{43}$ erg) is the total energy, $10^{-1} \theta_{-1}$ radians is the channeling angle, $10^{28} D_{28}$ cm (or $10^{24} D_{24}$ cm) is the luminosity distance, and $t_\gamma$ sec is the observer frame duration of the GRB event.  Given the uncertainty to which these quantities, the factor of proportionality in Equation 2, and $F_0 \sim 10^{-10}$ are known, the piston model cannot be ruled out, nor can it distinguish between a cosmological or a galactic halo GRB event.  The fact that the piston model is the simplest and most natural of the GRBR models makes its consistency with the reported GRB 970228 measurements particularly appealing.

However, at least two of the reported measurements appear to be inconsistent with the piston model.  
The first is the V-band upper bound of March 4 (van Paradijs et al. 1997), which is 1.0 magnitudes fainter than the best fit.  This is a difference of $\sim$ 3 times the expected photometric error.  
The second is the J-band measurement of March 17 (Klose et al. 1997), which is 1.3 magnitudes brighter than the best fit.  Klose et al. report that this is a 5-$\sigma$ detection, so it cannot easily be dismissed.
Within days of each of these measurements, optical measurements were taken that agree with the best fit to within the expected photometric error (Figure 1).  Consequently, the emission of the optical/infrared transient may be varying - both increasing and decreasing - by factors of $\sim$ 3 on timescales of days or weeks.  
If these measurements are correct, the nature of this emission would need to be explained and reconciled with the piston model before it can be fully accepted.
It should be noted that the 86.4 GHz upper bound of March 7 (Smith et al. 1997) and the 5 GHz upper bound of March 1 + 2 (Galama et al. 1997, Groot et al. 1997a) do not contradict the piston model because emission at these frequencies would be self-absorbed at these times.

Whereas the measurements taken before March 26 do not distinguish between cosmological and galactic halo distance scales, in terms of the piston model, the three ground-based optical and infrared measurements taken after March 26 can be used to determine whether or not the extended source is fading, which can distinguish between these distance scales for GRB 970228.  On March 26, the total emission of the extended object is approximately equal to that of the point source on that date.  
If the extended object is not fading, as would be expected for a galaxy, the R-band measurement of April 5 + 6 (Metzger et al. 1997b) and possibly the J- and K-band measurements of March 30 + 31 (Soifer et al. 1997) would be dominated by the emission of the extended object.  
However, all three of these measurements agree with the best fit of the transient point source to within their quoted photometric errors of $\pm 0.3$ and $\pm 0.2$ magnitudes, respectively.
Consequently, the extended object is probably fading.  If it were not, the R-band measurement would have been $\sim$ 0.8 magnitudes brighter.  
This difference estimate is given by adding the R-band flux of the extended object on March 26, based upon its approximate equality with that of the point source on this date, to the best-fit R-band flux of the point source on April 5 - 6.
Similar magnitude differences are found for the J- and K-band measurements, with the exact value of the difference depending on the color of the extended object.  
This conclusion is in agreement with the Keck II observations of Metzger et al. (1997a,b).  
They report that that the extended object, which is clearly visible in the R-band image of March 6 (Metzger et al. 1997a), is not observed to a deeper magnitude limit in the R-band image of April 5 + 6 (Metzger et al. 1997b).  
Fox et al. (1997), however, report that the extend object is consistent with not fading between the temporally closer HST observations of March 26 and April 7.  However, they additionally conclude that magnitude decreases of 0.48 (V-band) and 2.42 (I-band) cannot be ruled out to the 90\% confidence level.
Consequently, if the point source is indeed the dominant source of emission before March 26, then the extended object is probably fading and a galactic halo GRB event is favored by light travel time arguments.  This event may be described by the piston model, however the nature of the emission of the fading extended object must then be explained and reconciled with the piston model before it can be fully accepted.

\section{The Dominant Extended Source Scenario}

If the extended source is indeed fading, and if it is doing so more quickly than the point source is fading, then at some time before March 26, the extended source was likely the dominant source of the emission.  
Consequently, only the five optical and the two infrared measurements taken after and during March 26 can be used to analyze the emission of the point source in this case.  
The March 26 HST measurements are included because HST's resolution is sufficient to separate the emission of the point source from that of the extended object.  I fit Equation 1 to these seven measurements and find the following best-fit parameters values:  $\log F_0 = -14.0 \pm 5.9$, $a = 0.77 \pm 0.26$, and $b = 0.63 \pm 0.82$.  The standard deviation about the best fit is $\pm 0.19$ magnitudes, which is consistent with the expected standard photometric error.  
Once again, the 1-$\sigma$ errors of the best fit are only approximate.

The fitted value of the spectral parameter, $a = 0.77 \pm 0.26$, is again consistent with the synchrotron/inverse Compton values of standard GRBR models.
However, since the fitted fluxes are tightly clustered in $\log t$, the discriminatory parameters, $b$ and $F_0$, are not well constrained.  
If the temporal decay parameter, $b$, is much less than unity, the standard GRBR models are not applicable (\S2).
However, if $b \ga$ 1, the point source will dominate the emission at all times, which contradicts the underlying assumption of this scenario.  
Consequently, if the extended object dominates the emission at early times, before March 26, the point source may be compatible with the piston model, but not with the other standard GRBR models.  The nature of a fading and once dominant extended object would certainly need to be explained and reconciled with the piston model before it could be accepted.

\section{Conclusions}

In conclusion, the transient x-ray, optical, and infrared emission during the first 37 days following the GRB 970228 event, as well as the gamma ray, optical, infrared, millimeter, and radio upper bounds, appear to be consistent with that of a forward shocking impulsive fireball GRBR:  the piston model.  However, at least two discrepant measurements must be revisited, and if correct, they suggest that the post-burst flux of GRB 970228 may be varying by factors of $\sim$ 3 on timescales of days or weeks.  
This has yet to be reconciled with the piston model.  Furthermore, the post-March 26 ground-based optical and possibly infrared measurements suggest that the extended object has faded below the level at which it appeared in the HST images of March 26.  
If the extended object is indeed fading, then it is not a galaxy, and GRB 970228 may be at galactic halo distances.  
The piston model may still apply in this case, but the nature of a fading extended object would need to be explained and reconciled with the piston model before it could be fully accepted.

\acknowledgments
This research has been supported by NASA grant NAG5-2868.  I am grateful to D. Q. Lamb, B. P. Holden, F. J. Castander, D. M. Cole, and P. M\'esz\'aros for useful discussions.

\clearpage

\begin{deluxetable}{ccccc}
\footnotesize
\tablecolumns{6}
\tablewidth{0pc}
\tablecaption{Observations of the GRB 970228 Transient}
\tablehead{
\colhead{Band} & \colhead{Date (UT)\tablenotemark{a}} & \colhead{Flux/Magnitude} & \colhead{Telescope/Instrument} & \colhead{Reference}} 
\startdata
40 - 80 keV & F28.15 - F28.44 & $<$ 5.3 x 10$^{-5}$ ph cm$^{-2}$ s$^{-1}$ keV$^{-1}$ & CGRO/OSSE & Matz et al. 1997 \nl
40 - 80 keV & F28.15 - M3.00 & $<$ 2.2 x 10$^{-5}$ ph cm$^{-2}$ s$^{-1}$ keV$^{-1}$ & CGRO/OSSE & Matz et al. 1997 \nl
40 - 80 keV & F28.47 - F28.63 & $<$ 8.3 x 10$^{-5}$ ph cm$^{-2}$ s$^{-1}$ keV$^{-1}$ & CGRO/OSSE & Matz et al. 1997 \nl
2 - 10 keV & F28.45 & ($2.8 \pm 0.4$) x 10$^{-12}$ erg cm$^{-2}$ s$^{-1}$ & BeppoSAX/MECS & Costa et al. 1997b \nl
2 - 10 keV & M3.73 & $\sim$ 1.4 x 10$^{-13}$ erg cm$^{-2}$ s$^{-1}$ & BeppoSAX/MECS & Costa et al. 1997b \nl
2 - 10 keV & M7.03 - 7.49 & ($9.0 \pm 2.6$) x 10$^{-14}$ erg cm$^{-2}$ s$^{-1}\tablenotemark{b}$ & ASCA/GIS & Yoshida et al. 1997 \nl
2 - 10 keV & M7.03 - 7.49 & ($7.2 \pm 2.1$) x 10$^{-14}$ erg cm$^{-2}$ s$^{-1}\tablenotemark{b}$ & ASCA/SIS & Yoshida et al. 1997 \nl
0.5 - 10 keV & F28.46 & ($4.0 \pm 0.6$) x 10$^{-12}$ erg cm$^{-2}$ s$^{-1}$ & BeppoSAX/LECS & Costa et al. 1997b \nl
0.5 - 10 keV & M3.73 & $\sim$ 2.0 x 10$^{-13}$ erg cm$^{-2}$ s$^{-1}$ & BeppoSAX/LECS & Costa et al. 1997b \nl
0.1 - 2.4 keV & M10.79 - M13.32 & ($3.8 \pm 1.2$) x 10$^{-14}$ erg cm$^{-2}$ s$^{-1}\tablenotemark{c}$ & ROSAT/HRI & Frontera et al. 1997 \nl
B$_J$ & M3.1 & $23.3 \pm 0.5$ & ARC 3.5-m & Margon et al. 1997 \nl
B & M9.9 & 25.4 & INT & Groot et al. 1997b \nl
V & M1.0 & 21.3 & INT & Groot et al. 1997a \nl
V & M4.86 & $>$ 24.4 & NOT & van Paradijs et al. 1997 \nl
V & M8.9 & $>$ 23.6 & INT & Groot et al. 1997a \nl
V & M26.11 - M26.28 & $25.7 \pm 0.3$ & HST/WFPC2 & Sahu et al. 1997a\tablenotemark{d} \nl
V & A7.15 - A7.32 & $26.0 \pm 0.3$ & HST/WFPC2 & Sahu et al. 1997b\tablenotemark{d} \nl
R & M6.32 & 24.0 & Keck II & Metzger et al. 1997a\tablenotemark{d} \nl
R & M9.9 & 24.0 & INT & Groot et al. 1997b \nl
R & M11.18 & $\sim$ 24.0 & Palomar 5-m & Metzger et al. 1997a \nl
R & M13.0 & $23.8 \pm 0.2$ & NTT & Groot et al. 1997b\tablenotemark{d} \nl
R & A5.24 $+$ A6.27 & $24.9 \pm 0.3$ & Keck II & Metzger et al. 1997b \nl
$\sim$ 700 nm$\tablenotemark{e}$ & F28.81 & $m_K\tablenotemark{f}$ - ($1.6 \pm 0.5$) & RAO 0.9-m & Pedichini et al. 1997 \nl
$\sim$ 700 nm$\tablenotemark{e}$ & M4.81 & $> m_K\tablenotemark{f}$ + $\sim$ 1.1 & RAO 0.9-m & Pedichini et al. 1997 \nl   
I & M1.0 & 20.6 & WHT & Groot et al. 1997a \nl
I & M6.19 & 21.5\tablenotemark{g} & Palomar 1.5-m & Metzger et al. 1997a \nl
I & M8.9 & $>$ 22.2 & WHT & Groot et al. 1997a \nl
I & M26.11 - M26.28 & $24.2 \pm 0.3$ & HST/WFPC2 & Sahu et al. 1997a\tablenotemark{d} \nl
I & A7.15 - A7.32 & $24.6 \pm 0.3$ & HST/WFPC2 & Sahu et al. 1997b\tablenotemark{d} \nl
J & M17.8 & 21.0 & Calar Alto 3.5-m & Klose et al. 1997 \nl
J & M30.3 $+$ M31.2 & $23.5 \pm 0.2$ & Keck I & Soifer et al. 1997 \nl
H & M17.8 & $>$ 20.0 & Calar Alto 3.5-m & Klose et al. 1997 \nl
K & M17.8 & $>$ 19.5 & Calar Alto 3.5-m & Klose et al. 1997 \nl
K & M30.2 & $22.0 \pm 0.2$ & Keck I & Soifer et al. 1997 \nl
86.4 GHz & M7 & $<$ 1.2 mJy & BIMA & Smith et al. 1997 \nl
5 GHz & M1.75 $+$ M2.75 & $<$ 0.35 mJy & WSRT & Groot et al. 1997a \nl
\enddata
\tablenotetext{a}{1997 February 28.15 - 1997 April 7.32}
\tablenotetext{b}{Assumes a power law spectrum of photon index 1.4}
\tablenotetext{c}{Assumes a power law spectrum of photon index 1.9}
\tablenotetext{d}{Extended object detected}
\tablenotetext{e}{FWHM $\sim$ 300 nm}
\tablenotetext{f}{Magnitude of nearby K star (R $\sim$ 21.5 (Groot et al. 1997b), 22.4 (Metzger et al. 1997a))}
\tablenotetext{g}{Measured near detection threshold}
\end{deluxetable}

\clearpage



\clearpage

\figcaption[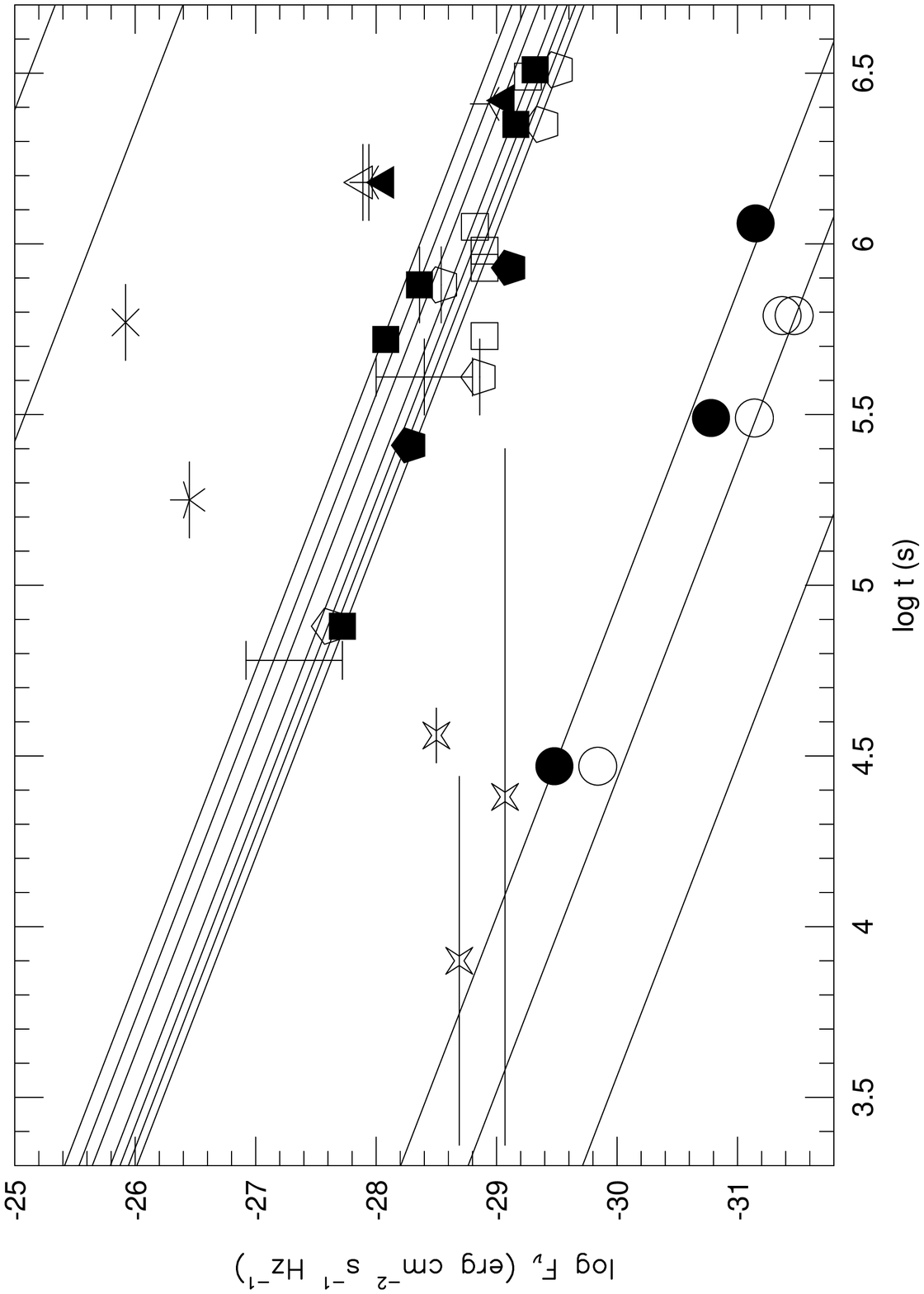]{Fluxes and flux upper bounds for the observations of the GRB 970228 transient (Table 1) and the best fit to Equation 1.  Fluxes (symbols) and flux upper bounds (symbols with horizontal lines) have been corrected for galactic extinction as described in \S2.1.  Stars are the 40 - 80 keV gamma ray band, open circles are the 2 - 10 keV x-ray band, solid circles are the 0.5 - 2 keV x-ray band, solid pentagons are B-band, open pentagons are V-band, open squares are R-band, solid squares are I-band, solid triangles are J-band, open triangles are H-band, three-prongs are K-band, four prongs are the 86.4 GHz millimeter band, and five-prongs are the 5 GHz radio band.  
The large error bars represent the broadband measurements of Pedichini et al. (1997).  
The lowest line is the best fit temporal decay for the 40 - 80 keV gamma ray band, the second lowest line is that for the 2 - 10 keV x-ray band, etc.\label{r97fig.ps}}

\clearpage 

\setcounter{figure}{0}
\begin{figure}[tb]
\plotone{r97fig.ps}
\end{figure}

\end{document}